\documentclass{ws-p10x7}
\usepackage{hyperref}
\def\lesssim{\mathrel{\mathpalette\vereq<}}
\def\gtrsim{\mathrel{\mathpalette\vereq>}}
\makeatletter
\def\vereq#1#2{\lower3pt\vbox{\baselineskip1.5pt \lineskip1.5pt
\ialign{$\m@th#1\hfill##\hfil$\crcr#2\crcr\sim\crcr}}}
\makeatother

\title{Theory of Neutrino Masses and Mixings}

\author{Hitoshi Murayama}

\address{Department of Physics, University of California\\
  Berkeley, CA 94720, USA}

\address{Theoretical Physics Group, Lawrence Berkeley National
  Laboratory\\
  University of California, Berkeley, CA 94720, USA\\
  E-mail: \url{murayama@lbl.gov}}

\begin{document}

\twocolumn[\maketitle\abstract{ Neutrino physics is going through a
  revolutionary progress.  In this talk I review what we have learned
  and why neutrino mass is so important.  Neutrino masses and mixings
  are already shedding new insight into the origin of flavor.  Given
  the evidences for neutrino mass, leptogenesis is gaining momentum as
  the origin of cosmic baryon asymmetry.  Best of all, we will learn a
  lot more in the coming years.  }]

\section{Introduction}

There is no question that we are in a truly revolutionary moment in
neutrino physics.  From the previous speakers, Jordan
Goodman,\cite{Goodman} Josh Klein,\cite{Klein} Chang-Kee
Jung,\cite{Jung} and Shigeki Aoki,\cite{Aoki} we have seen a wealth of
new experimental data in neutrino physics since the previous Lepton
Photon conference.  We are going through a revolution in our
understanding of neutrinos, even in broader context of flavor physics,
unification, and cosmology.  My job is to go through some of the
exciting aspects of these topics.

I found that not only us but also general public is excited about
neutrino physics; I've found fortune cookies from SuperK brand on my
way from Snowmass meeting to Denver airport at a Chinese restaurant.

\begin{figure}[t]
  \begin{center}
    \includegraphics[width=\columnwidth]{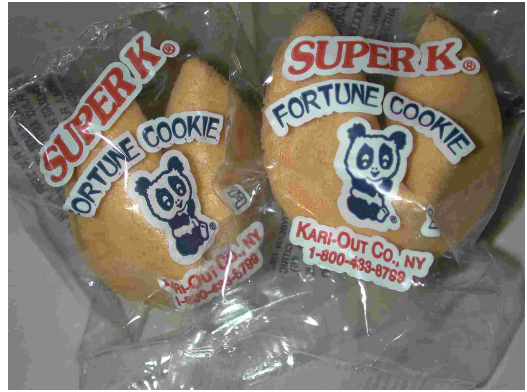}
    \caption{Fortune cookies from SuperK brand I found in Colorado.}
    \label{fig:fortune}
  \end{center}
\end{figure}

We have learned the following important points since the previous 
Lepton Photon conference:
\begin{itemize}
\item Evidence for $\nu_{\mu}$ deficit in atmospheric neutrinos is
  stronger than ever, with the up/down asymmetry established at more
  than 10~$\sigma$ level in $\nu_{\mu}$-induced events.\cite{Goodman}
  We are more than 99\% certain $\nu_{\mu}$ are converted mostly to
  $\nu_{\tau}$.  Current K2K data support this evidence.\cite{Jung}
    
\item Putting SuperKamiokande and SNO data together, we are certain at
  3~$\sigma$ level that the solar $\nu_{e}$ must have converted to
  $\nu_{\mu}$ or $\nu_{\tau}$.\cite{Klein}
    
\item Such neutrino conversions are most likely due to neutrino
  oscillations.  Other possibilities, such as neutrino decay,
  violation of equivalence principle, spin-resonant rotation, exotic
  flavor-changing interactions, are still possible, but are either
  squeezed phenomenologically or rely on models that are theoretically
  not motivated or esthetically not pleasing.
    
\item Tiny neutrinos masses required in neutrino oscillation for
  atmospheric and solar neutrinos are the first evidence for {\it the
    incompleteness of the Minimal Standard Model.}!  Even apart from
  oscillations, any explanation to these phenomena require physics
  beyond the Standard Model.
\end{itemize}

Given this dramatic progress in experimental results, it is
interesting to see how good insight theorists had had on neutrino
physics.  Here is the list of typical views among theorists on neutrino
physics back in 1990:
\begin{itemize}
\item Solar neutrino problem must be solved by the small angle MSW 
  solution because it is so cute.
  
\item The natural scale for $\nu_{\mu}$-$\nu_{\tau}$ oscillation 
  is $\Delta m^{2} = 10$--100~eV$^{2}$ because it is cosmologically 
  interesting.
  
\item The mixing angle for $\nu_{\mu}$-$\nu_{\tau}$ oscillation is 
  of the order of $V_{cb}$.
  
\item Atmospheric neutrino anomaly must go away because it
  requires an ugly large angle, not suggested by simple
  grand-unified theories.
\end{itemize}    
Looking back at this list, the first one is most likely wrong as we 
will see later, and (2--4) are all wrong.  You can see that theorists 
had had great insight into the nature of neutrinos.

The rest of the talk is organized in the following manner.  I first
review the situation with the global fit to solar neutrino data after
SNO results came out.  Then models of neutrino masses and mixings, in
the broader context of models of flavor, are briefly reviewed.  After
that, the idea of leptogenesis is explained, which is one of the main
ideas now to explain the cosmological matter anti-matter asymmetry. 
Finally I will discuss what we can look forward to in the near future.

\section{Global Fits}

I review the global fits to the solar neutrino data including SNO.
But before doing so, I spend some time discussing the parameter space
of two-flavor neutrino oscillation.

Traditionally, neutrino oscillation data had been shown on the
$(\sin^{2} 2\theta, \Delta m^{2})$ plane.  This parameterization,
however, covers only a half of the parameter space.  We instead need
to use, for example, $(\tan^{2} \theta, \Delta m^{2})$ to present
data.  This point had been recognized for a long time in the context
of three-flavor mixing,\cite{Fogli:1996ne} but two-flavor
analyses had always been presented on the $(\sin^{2} 2\theta, \Delta
m^{2})$ except as a limit of three-flavor analyses.  Because you
will see plots that cover the full parameter space soon and also in
the future, I will briefly explain this point.

The oscillation occurs because the flavor eigenstate ({\it i.e.}\/ 
$SU(2)_{L}$ partner of charged leptons) and the mass eigenstates are 
not the same.  Let us talk about $\nu_{e}$ and $\nu_{\mu}$ for the 
sake of discussion.  The mass eigenstates are then given by two
orthogonal linear combination of them:
\begin{eqnarray}
    \nu_{1} &=& \nu_{e} \cos\theta + \nu_{\mu} \sin\theta, \\
    \nu_{2} &=& -\nu_{e} \sin\theta + \nu_{\mu}\cos\theta.
\end{eqnarray}
As a convention, we can always choose $\nu_{2}$ to be heavier than
$\nu_{1}$ without a loss of generality.  Now the question is how much
we should vary $\theta$.  If you use $\sin^2 2\theta \in [0,1]$ as
your parameter, $\theta$ can go from 0 to $45^\circ$ to exhaust all
possibilities of $\sin^2 2\theta$.  However, for $\theta\leq 45^\circ$,
$\nu_1$ always contains more $\nu_e$ than $\nu_\mu$.  To represent the
possibility of $\nu_1$ dominated by $\nu_\mu$ rather than $\nu_e$, we
need to allow $\theta$ to go beyond $45^\circ$ up to $90^\circ$.  Then
$\sin^2 2\theta$ folds over at 1 and comes back down to 0.  Clearly,
$\sin^2 2\theta$ is not the right parameter for global fits to the
data.  

When the oscillation is purely that in the vacuum (no matter effect),
the oscillation probability depends only on $\sin^2 2\theta$, 
\begin{equation}
  P(\nu_e \rightarrow \nu_\mu) = \sin^2 2\theta \sin^2 1.27
  \frac{\Delta m^2}{E} L,
\end{equation}
with $\Delta m^2$ in eV$^2$, $E$ in GeV and $L$ in km.  Therefore, it
is the same for $\theta$ and $90^\circ -\theta$, explaining why
$\sin^2 2\theta$ had been used for fits in the past.  In the presence
of the matter effect, however, the oscillation probability is
different for $\theta$ and $90^\circ -\theta$.  

To cover all physically distinct possibilities, one possible parameter
choice is $\sin^2 \theta$, that ranges from 0 to 1 for $\theta \in
[0^\circ, 90^\circ]$, and shows the symmetry under $\theta
\leftrightarrow 90^\circ - \theta$ in the absence of the matter effect
as a reflection with respect to the axis $\theta = 45^\circ$ on a
linear scale.  This is a perfectly adequate choice for representing
atmospheric neutrino data, for instance when matter effect is included
in $\nu_\mu \leftrightarrow \nu_s$ oscillation.  On a log-scale,
however, $\tan^2\theta$ shows the symmetry manifestly as $\theta
\rightarrow 90^\circ - \theta$ takes $\tan^2 \theta \rightarrow \cot^2
\theta = 1/\tan^2 \theta$.  Especially for fits to the solar neutrino
data, we need to cover a wide range of mixing angles, and hence a
log-scale; that makes $\tan^2 \theta$ essentially the unique choice
for graphically presenting the fits.  The range of mixing angle
$0^\circ \leq \theta \leq 45^\circ$ is what had been covered
traditionally with the parameter $\sin^2 2\theta$, and we call it
``the light side,'' while the remaining range $45^\circ \leq \theta
\leq 90^\circ$ ``the dark side'' because it had been usually neglected
in the fits.\cite{deGouvea:2000cq}

Now that SNO charged-current data is available, we would like to see
its impact on the global fit to the solar neutrino data.  Of course,
the most important lesson from the SNO data is that there is an
additional active neutrino component $\nu_a$ ({\it i.e.}\/, a linear
combination of $\nu_\mu$ and $\nu_\tau$) coming from the Sun.  The
next quantitative lesson is that BP00 flux calculation of $^8$B
neutrino flux is verified within its error, as discussed by Josh Klein
in his talk.\cite{Klein}

\begin{figure}[t]
  \begin{center}
    \includegraphics[width=\columnwidth]{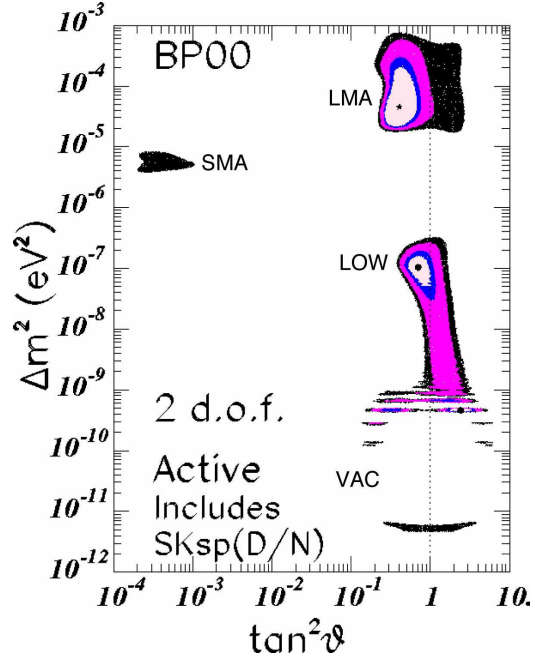}
    \caption[twodof]{Two-flavor fit to solar neutrino data for $\nu_e$ to
      $\nu_a$ oscillation\cite{Bahcall:2001zu}. The confidence levels
      are 90\%, 95\%, 99\%, and 99.73\% (3$\sigma$).}
    \label{fig:twodof}
  \end{center}
\end{figure}

I show results from the global fits.\cite{Bahcall:2001zu} The first
one is a two-flavor fits for $\nu_e$ to $\nu_a$ oscillation, using
BP2000 solar neutrino flux calculations in Fig.~\ref{fig:twodof}.
Four regions of the parameter space, LMA (Large Mixing Angle MSW
solution), SMA (Small Mixing Angle MSW solution), LOW (MSW solution
with LOW $\Delta m^2$), and VAC (VACuum oscillation solution), that
fit the data can be seen.  LMA solution is the best fit to the current
data.

\begin{figure}[t]
  \begin{center}
    \includegraphics[width=\columnwidth]{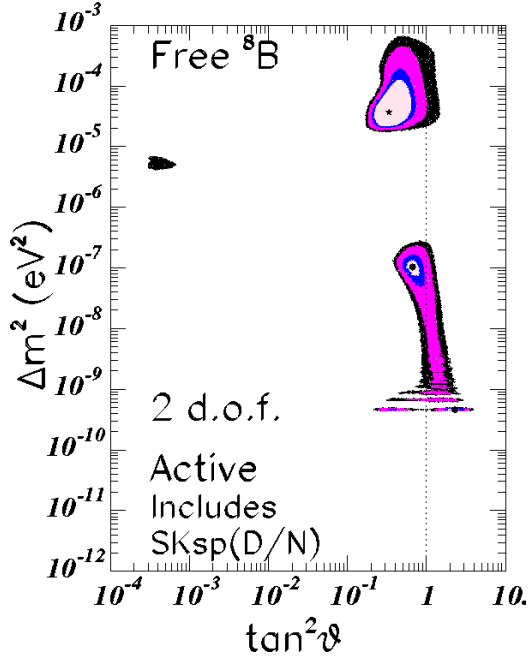}
    \caption[twodoffree]{Two-flavor fit taking the solar $^8$B neutrino flux
      as a free parameter.\cite{Bahcall:2001zu}}
    \label{fig:twodoffree}
  \end{center}
\end{figure}

One of the concerns in solar neutrino fits has been that been that
solar $^8$B neutrino flux is very sensitive to solar
parameters.\footnote{This concern has been greatly ameliorated by the
  agreement of helioseismology data and the standard solar model,
  however.\cite{helio}}  Thanks to the SNO data, we can now drop the predicted
flux entirely from the fit and obtain equally good fit, as shown in
Fig.~\ref{fig:twodoffree}.

\begin{figure}[t]
  \begin{center}
    \includegraphics[width=0.5\columnwidth]{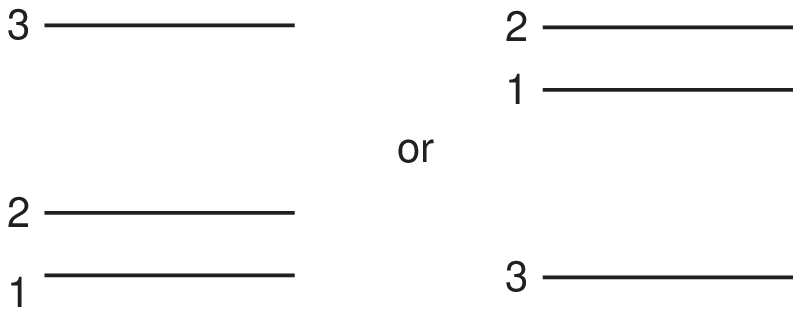}
    \caption{The mass spectrum of three neutrino mass eigenstates.}
    \label{fig:2+1}
  \end{center}
\end{figure}

To account for both solar and atmospheric neutrino oscillations, we
need three-flavor analyses.  The standard parameterization of the
mixing among neutrinos, MNS (Maki--Nakagawa--Sakata) matrix, is given
by
\begin{eqnarray}
  \lefteqn{
  U_{MNS} = \left( \begin{array}{ccc} U_{e1} & U_{e2} & U_{e3} \\
      U_{\mu 1} & U_{\mu 2} & U_{\mu 3} \\
      U_{\tau 1} & U_{\tau 2} & U_{\tau 3} 
    \end{array} \right) } \nonumber \\
  &=& \left( \begin{array}{ccc} 1 & 0 & 0\\
      0 & c_{23} & s_{23} \\ 0 & -s_{23} & c_{23} 
    \end{array} \right)
  \left( \begin{array}{ccc} c_{13} & 0 & s_{13} e^{-i\delta} \\
      0 & 1 & 0\\
      - s_{13} e^{i\delta} & 0 & c_{13} 
    \end{array}\right) \nonumber \\
  & & \times
  \left( \begin{array}{ccc} c_{12} & s_{12} & 0\\
      -s_{12} & c_{12} & 0 \\
      0 & 0 & 1
    \end{array} \right).
  \label{eq:UMNS}
\end{eqnarray}
The angle $\theta_{13}$ is currently undetermined, except that there
is an upper bound $\sin \theta_{13} \lesssim 0.16$ from reactor
neutrino experiments\cite{CHOOZ} for $\Delta m^2 = 3 \times
10^{-3}~\mbox{eV}^2$ preferred by the atmospheric neutrino data.
Fortunately, the smallness of $\theta_{13}$ essentially decouples
solar and atmospheric neutrino oscillation, allowing us to interpret
data with separate two-flavor fits.  The convention here is that the
mass eigenstates $\nu_1$ and $\nu_2$ have small mass splitting for the
solar neutrino oscillation $\Delta m^2_{12} \lesssim 2 \times 10^{-4}
\mbox{eV}^2$, while $\nu_2$ and $\nu_3$ have the splitting $\Delta
m^2_{23} \simeq 3 \times 10^{-3}~\mbox{eV}^2$, as shown in
Fig.~\ref{fig:2+1}.  I've put $\nu_2$ above $\nu_1$ assuming we are on
the ``light side'' of solar neutrino oscillation.  The ``dark side''
of solar neutrino oscillation would correspond to the opposite order.
Future improvements in the solar neutrino data would allows us to
discriminate between the two.  In addition, we still do not know if
$\nu_3$ should be above or below the solar doublet.  The remaining
issues are (1) to determine the solar neutrino oscillation parameters,
(2) the ordering of mass eigenstates, and (3) $\theta_{13}$.

\begin{figure}[t]
  \begin{center}
    \includegraphics[width=\columnwidth]{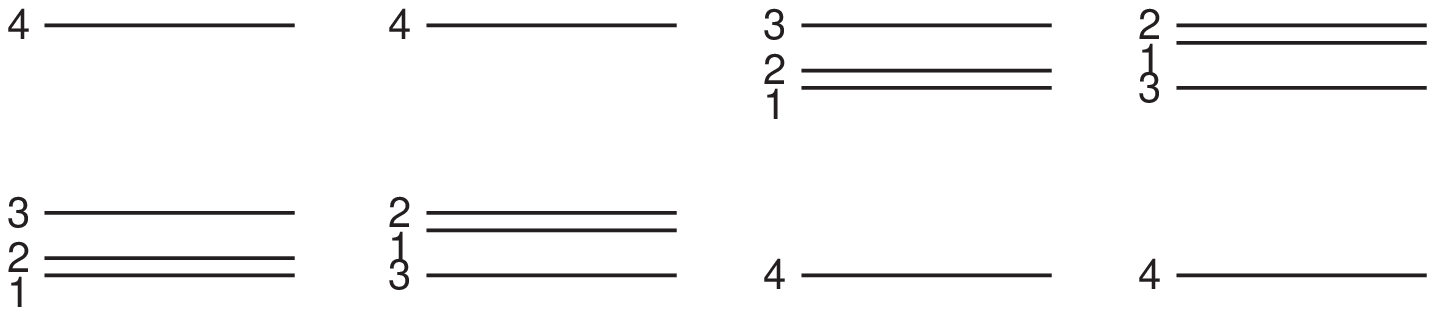}
    \caption{The 3+1 mass spectra of four neutrino mass eigenstates.}
    \label{fig:3+1}
    \includegraphics[width=0.5\columnwidth]{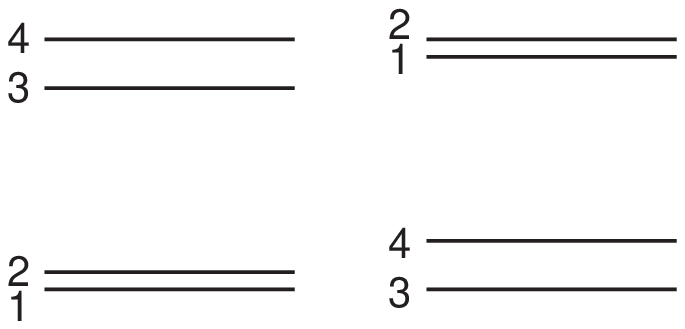}
    \caption{The 2+2 mass spectra of four neutrino mass eigenstates.}
    \label{fig:2+2}
  \end{center}
\end{figure}

When LSND oscillation signal (see talk by Aoki\cite{Aoki}), that
prefers $\Delta m^2 \sim \mbox{eV}^2$, is also considered, we have to
accommodate three different orders of magnitude of $\Delta m^2$ values
and hence four mass eigenstates.  Because of three $\Delta m^2$, the
mass spectrum now has $3!=6$ possibilities shown in
Figs.~\ref{fig:3+1} and \ref{fig:2+2}.

\begin{figure*}[t]
  \begin{center}
    \includegraphics[width=2\columnwidth]{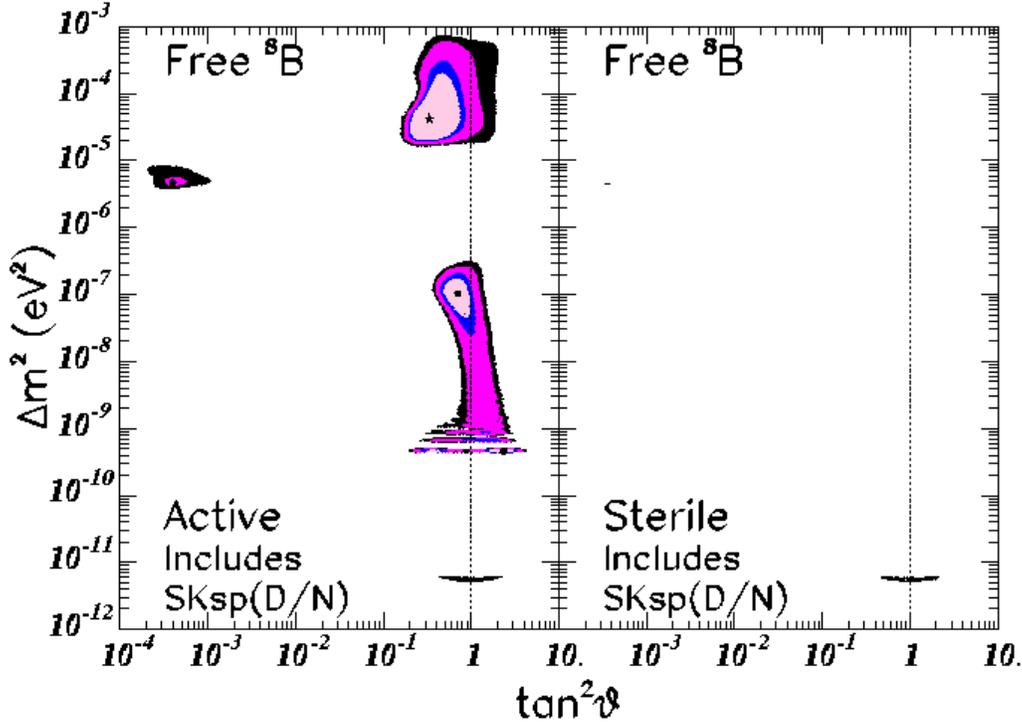}
    \caption[twodoffree]{A global fit interpolating the pure active
      and pure sterile cases.\cite{Bahcall:2001zu}  See text for the
      precautions on the definition of confidence levels that differ
      from Figs.~\ref{fig:twodof} and \ref{fig:twodoffree}.}
    \label{fig:glo_acst_bo_sno}
  \end{center}
\end{figure*}

Now that both solar and atmospheric neutrino data disfavor
oscillations into pure sterile state, the scenario with sterile
neutrino is getting squeezed.  For example, the comparison of sterile
and active case can be done by interpolating the two cases with an
additional parameter $\eta$.  The state that solar $\nu_e$ oscillates
to is a mixture of an active and sterile neutrino, $\nu_a \cos\eta +
\nu_s \sin\eta$.  Then one can fit the data with three degrees of
freedom, $\Delta m^2$, $\tan^2\theta$, and $\eta$, and finds that the
fit is better for $\eta = 0$ (pure active case).  A word of caution
here is that the definition of the confidence levels is now looser,
making the allowed regions bigger.  The confidence levels 90\%, 95\%,
99\%, and 99.73\% in Fig.~\ref{fig:glo_acst_bo_sno} would correspond
to 96\%, 98\%, 99.7\%, and 99.92\% for pure two-flavor case.
Nonetheless, it is clear that the pure sterile case provides a much
less good fit.

\begin{figure*}[t]
  \begin{center}
    \includegraphics[width=2\columnwidth]{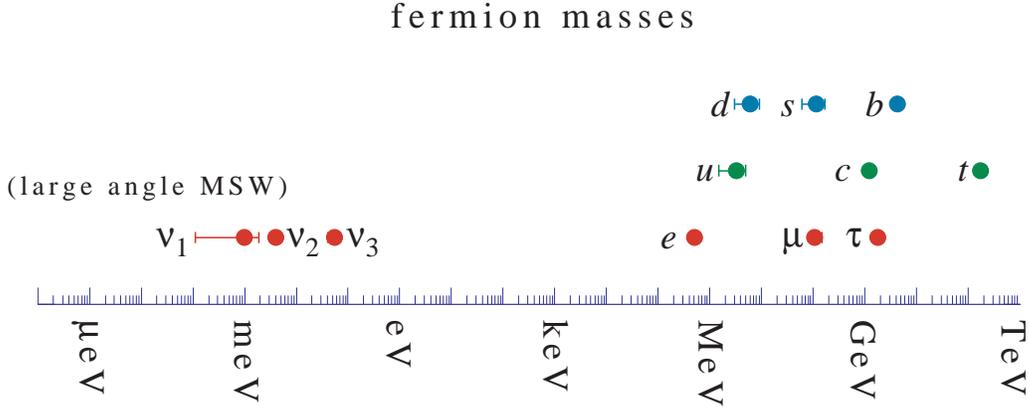}
    \caption[masses]{Mass spectrum of quarks and leptons.  LMA
      solution to the solar neutrino problem is assumed, while the
      range given for $\nu_1$ mass is basically just a guess.} 
    \label{fig:masses}
  \end{center}
\end{figure*}

However, the phenomenological motivation for having a sterile state
comes from the combination of LSND, atmospheric, and solar neutrino
data, and it requires full four-flavor analysis.  In the case of $3+1$
spectra, the mixing angle for LSND oscillation is related to the MNS
matrix elements as $\sin^2 2\theta_{\rm LSND} = 4 |U_{e4}|^2 |U_{\mu
  4}|^2 \gtrsim 0.01$.  On the other hand, $|U_{\mu 4}|$ cannot be too
large because of CDHS data and atmospheric neutrino data, while
$|U_{e4}|^2$ either because of Bugey reactor neutrino
data.\cite{Bilenkii:1999ny} This type of spectra is only marginally
allowed\cite{Barger:2000ch}, and fits prefer 2+2
spectra.\cite{Maltoni:2001mt} However, 2+2 spectra are also getting
squeezed now that both solar and atmospheric neutrino data disfavor
oscillations into pure sterile state.  If solar neutrino oscillation
is into a pure active state, the atmospheric neutrino oscillation must
be into a pure sterile state, and vice versa. The way this type of
spectra can fit the data is to find a compromise between the
requirements of sufficient active component in both oscillations.
Detailed numerical analysis showed that such a compromise is still
possible and the sterile neutrino is still
allowed.\cite{Gonzalez-Garcia:2001zi} The final verdict will be given
by Mini-BooNE as we will see later.\footnote{There is an intriguing
  possibility that all three oscillation data can be explained without
  a sterile neutrino, if there is CPT violation that allows different
  mass spectra between neutrinos and
  anti-neutrinos.\cite{Murayama:2000hm} This possibility can be tested
  by having anti-neutrino run at Mini-BooNE beyond the planned
  neutrino run.}

\section{Models}

Now that neutrinos appear massive, despite what Standard Model has
assumed for decades, we need to somehow incorporate the neutrino
masses into our theory.  The most striking facts about the neutrino
sector are (1) the masses are very small, and (2) mixing angles appear
large.  Looking at the mass spectrum of quarks and leptons, it is
especially bizarre that even the third-generation mass is so low
compared to quarks and charged leptons.  An explanation is clearly
called for.

The minute you talk about masses of spin 1/2 particles, you need both
spin up and down states because you can stop any massive particle.
When the particle is at a relativistic speed, a more useful label is
left- or right-handed states.  For strictly massless particles, left-
and right-handed states are completely independent from each other and
you do not need both of them; this is how neutrinos are described in
the Standard Model.  Once they are massive, though, we need both, so
that we can write a mass term using both of them:
\begin{equation}
  {\cal L}_{\rm mass} = m_D (\overline{\nu_L} \nu_R + \overline{\nu_R}
  \nu_L).
  \label{eq:mass}
\end{equation}
But then the mass term is exactly the same as the other quarks and
leptons, and why are neutrinos so much lighter?

\begin{figure}[t]
  \begin{center}
    \includegraphics[width=0.5\columnwidth]{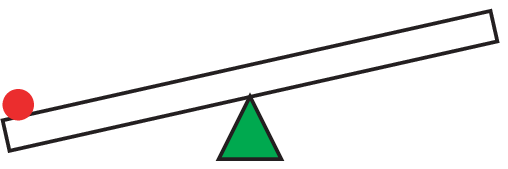}
    \caption{Too large Dirac mass of neutrinos.}
    \label{fig:seesaw1}
    \includegraphics[width=0.5\columnwidth]{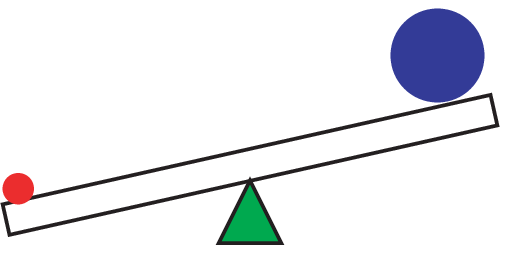}
    \caption{We can put a large mass to the right-handed neutrino
      without violating gauge invariance.}
    \label{fig:seesaw2}
    \includegraphics[width=0.5\columnwidth]{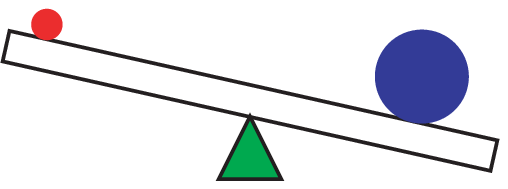}
    \caption{Then the mass of the neutrino becomes light.}
    \label{fig:seesaw3}
  \end{center}
\end{figure}

The so-called seesaw mechanism\cite{seesaw} is probably the most
motivated explanation to the smallness of neutrino
masses.\footnote{Recently, alternative explanation using extra
  dimensions had appeared.\cite{extraD}} The first step is to rewrite
the mass term Eq.~(\ref{eq:mass}) in a matrix form
\begin{equation}
  {\cal L}_{\rm mass} = \frac{1}{2}(\nu_L \ \overline{\nu_R})
  \left( 
    \begin{array}{cc}
      0 & m_D \\ m_D & 0
    \end{array}
  \right)
  \left(
    \begin{array}{c}
      \nu_L \\ \overline{\nu_R}
    \end{array}
  \right) + c.c.
\end{equation}
Here, I had to put $\nu_L$ and $\overline{\nu_R}$ (CP conjugate of
$\nu_R$) together so that both of them are left-handed and are allowed
to be in the same multiplet. The problem was that we (at least
naively) expect the ``Dirac mass'' $m_D$ to be of the same order of
magnitudes as other quarks and lepton masses in the same generation
which would be way too large (Fig.~\ref{fig:seesaw1}).  The point is
that the right-handed neutrino is completely neutral under the
standard-model gauge groups and is not tied to the electroweak
symmetry breaking ($v = 246$~GeV) to acquire a mass.  Therefore, it
can have a mass much larger than the electroweak scale without
violating gauge invariance, and the mass term is
(Fig.~\ref{fig:seesaw2})
\begin{equation}
  {\cal L}_{\rm mass} = \frac{1}{2}(\nu_L \ \overline{\nu_R})
  \left( 
    \begin{array}{cc}
      0 & m_D \\ m_D & M
    \end{array}
  \right)
  \left(
    \begin{array}{c}
      \nu_L \\ \overline{\nu_R}
    \end{array}
  \right) + c.c..
\end{equation}
Because one of the mass eigenvalues is clearly dominated by $M \gg
m_D$, while the determinant is $-m_D^2$, the other eigenvalue must be
suppressed, $- m_D^2/M \ll m_D$ (Fig.~\ref{fig:seesaw3}).  This way,
physics at high-energy scale $M$ suppresses the neutrino mass in a
natural way.  In order to obtain the mass scale for the atmospheric
neutrino oscillation $(\Delta m^2_{\rm atm})^{1/2} \sim 0.05$~eV, and
taking the third generation mass $m_D \sim m_t \sim 170$~GeV, we find
$M = m_D^2/m_\nu \sim 0.6 \times 10^{15}$~GeV.  It is almost the
grand-unification scale $2 \times 10^{16}$~GeV where all gauge
coupling constants appear to unify in th minimal supersymmetric
standard model.

Indeed, the seesaw mechanism was motivated by $SO(10)$ grand-unified
models which include right-handed neutrinos automatically together
with all other quarks and leptons in irreducible 16-dimensional
multiplets.  $SO(10)$ GUT also has an esthetic appeal, being the
smallest anomaly-free gauge group with chiral fermions, while not
requiring additional fermions beyond the right-handed neutrinos.  But
it has a slight problem: it is too predictive.  The simplest version
of the model predicts $m_c = m_s = m_\mu$ at the GUT-scale and no CKM
(Cabibbo--Kobayashi--Maskawa) mixing.  The art of unified model
building is how to break the naive prediction to a realistic one.

There are many many models, with or without grand unification, of
neutrino masses and mixings.  There are papers submitted by C.S. Lam,
Alexandre Khodin, Joe Sato, Koichi Matsuda, Bruce McKellar, and Carl
H.~Albright to this conference.  The bottom-line is that one can
construct nice unified models of quark and lepton masses, especially
within $SU(5)$ or $SO(10)$ unification.  I do not intend to go into an
exhaustive review of proposed models because I can't.\footnote{I hope
  no list angers less people than an incomplete list.}  Instead, I'd
like to discuss how we might eventually arrive at understanding masses
and mixings in a bottom-up approach.

Looking at quarks, the masses are very hierarchical, and mixings are
small.  The masses of charged leptons are also very hierarchical.
Even though we are used to hierarchical mass spectrum, it is actually
quite bizarre: why do particles with exactly the same quantum numbers
have such different masses, and mix little?  In graduate quantum
mechanics, didn't we learn that states with same quantum numbers have
typically similar energy levels and mix substantially?  A very naive
answer to the puzzle is that different generations of quarks
presumably have different hidden quantum numbers we have not
identified yet.  We call them ``flavor quantum numbers.''  Flavor
quantum numbers distinguish different generations, allowing them to
have very different energy levels (masses) and forbidding them to mix
substantially.

First question we should ask then is if there is need for fundamental
distinction among three neutrinos.  If you look at the
currently-favored LMA solution to the solar neutrino problem, two
mass-squared differences are not that different, $\Delta m^2_{\rm atm}
\simeq 1\mbox{--}7 \times 10^{-3}\mbox{eV}^2$, $\Delta m^2_{\rm LMA}
\simeq 0.2\mbox{--}6 \times 10^{-4}\mbox{eV}^2$ (both at 99\% CL).
The mixing angles $\theta_{12}$, $\theta_{23}$ are both large.  Even
though $\theta_{13}$ is usually said to be small, $|U_{e3}| = \sin
\theta_{13} \lesssim 0.16$ at $\Delta m^2_{\rm atm} = 3 \times
10^{-3}\mbox{eV}^2$, it is smaller than $|U_{\mu 3}| \simeq |U_{\tau
  3}| \simeq 1/\sqrt{2} = 0.71$ only by a factor of 2.3.  Furthermore
at the low end of $\Delta m^2_{\rm atm}$, $\sin \theta_{13}$ may be
still sizable (0.2 (0.4) at $\Delta m^2_{\rm atm} = 2 (1) \times
10^{-3}\mbox{eV}^2$).  It is not clear yet if $\theta_{13}$ is so
small. If there is no hierarchy and mixing is large, we apparently do
not need new quantum numbers to distinguish three generation of
neutrinos.

But isn't near-maximal mixing suggested by atmospheric neutrino data
special?  Isn't there a special reason for it?  If there is no
distinction among three generations of neutrinos, isn't even a small
hierarchy between atmospheric and solar mass-squared differences
puzzling?  To address these questions, we ran Monte Carlo over seesaw
mass matrices.\cite{anarchy} It turns out that the distribution is
peaked at $\sin^2 2\theta = 1$ for atmospheric neutrino mixing, and
the ratio of two masses-squared differences $\Delta m^2_{\rm
  solar}/\Delta m^2_{\rm atm} \sim 0.1$ is the most likely value.  The
lesson here is that, if there is no fundamental distinction among
three generations of neutrinos, the apparent pattern of masses and
mixings comes out quite naturally.  Of course, ``randomness'' behind
Monte Carlo is just a measure of our ignorance.  But complicated
unknown dynamics of flavor at some high-energy scale may well appear
to produce random numbers in the low-energy theory.  We call such a
situation ``anarchy.''  And the peak in the mixing angle distribution
can be understood in terms of simple group theory.\cite{anarchy2} In
fact, the anarchy predicts that all three mixings angles are peaked in
$\sin^2 2\theta$ distributions at maximum.  This does not sound quite
right for $\sin^2 2\theta_{13}$, but we find that three out of four
distributions, $\theta_{12}$, $\theta_{23}$, $\Delta m^2_{\rm
  solar}/\Delta m^2_{\rm atm}$, prefer what data suggest.  I find it
quite reasonable.  If you take this idea seriously, then $\sin^2
2\theta_{13}$ must be basically just below the current limit and we
hope to see it sometime soon.

\begin{figure}[t]
  \begin{center}
    \includegraphics[width=\columnwidth]{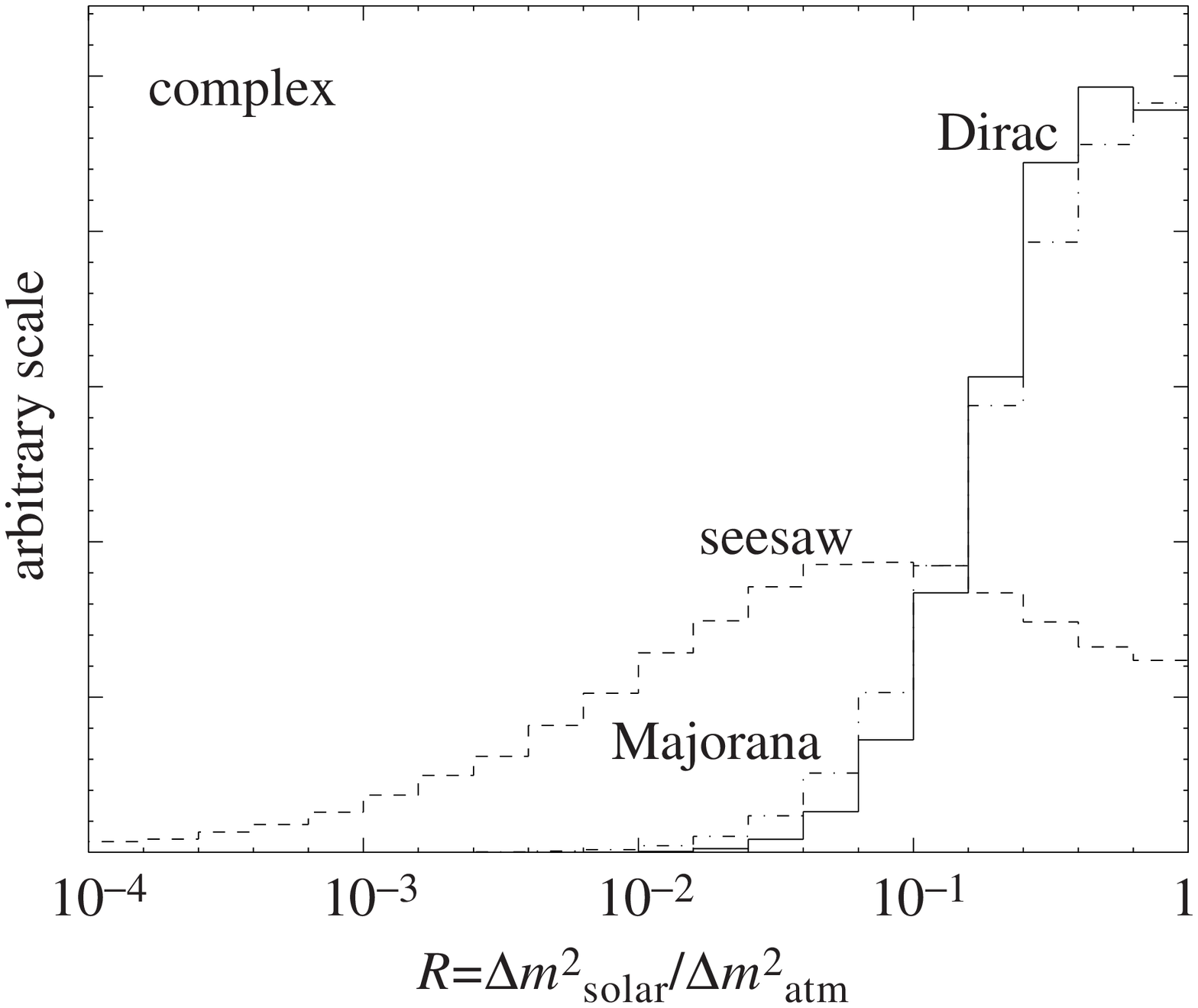}
    \caption[R]{The ratio of two mass-squared differences in randomly
      generated $3\times 3$ neutrino mass matrices.\cite{anarchy2}
      For the seesaw case, the peak is around $\Delta m^2_{\rm
        solar}/\Delta m^2_{\rm atm} \sim 0.1$.}
    \label{fig:R}
    \includegraphics[width=\columnwidth]{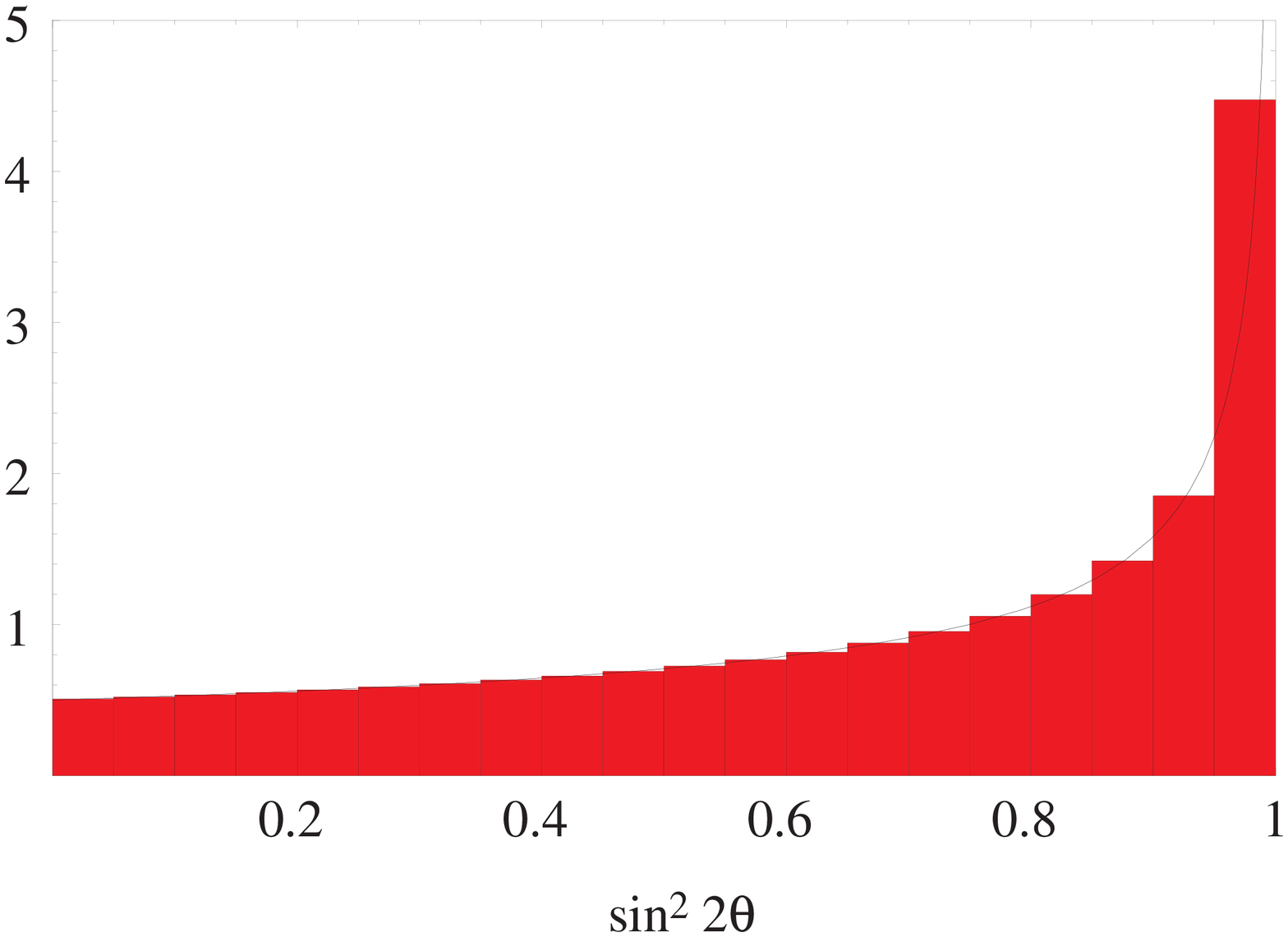}
    \caption[R]{$\sin^2 2 \theta_{23}$ in randomly
      generated $3\times 3$ neutrino seesaw mass matrices.\cite{anarchy2}}
    \label{fig:s23c}
  \end{center}
\end{figure}

What about quarks and charged leptons?  Here we clearly see a need for
fundamental distinction among three generations.\footnote{See also
  Riccardo Barbieri at this conference.\cite{Barbieri}}  They are
definitely not anarchical; they needed ordered hierarchical structure.
Let us suppose the difference among three generations is just a new
charge, namely a flavor $U(1)$ quantum number.  As a simple exercise,
we can assign the following flavor charges consistent with
$SU(5)$-type unification:
\begin{eqnarray}
  \label{eq:charges}
  & & {\bf 10} (Q, \bar{U}, \bar{E}) (+2, +1, 0),\nonumber \\
  & &{\bf 5}^* (L, \bar{D}) (+1, +1, +1).
\end{eqnarray}
All three generations of $L$ have the same charge because of anarchy:
no fundamental distinction among them.  As we saw, neutrino masses and
mixings come out reasonably well from this charge assignment.  With
$SU(5)$-like unification, right-handed down quarks, that belong to the
same ${\bf 5}^*$ multiplets with left-handed lepton doublets, also
have the same charge for all three generations.  It is intriguing that
large mixing among right-handed quarks is consistent with what we know
about quarks, because the CKM matrix is sensitive only to particles
that participate in charged-current weak interaction, namely
left-handed quarks.  Even if right-handed quarks are maximally mixed,
we wouldn't know.  On the other hand, ${\bf 10}$ multiplets, that
contain left-handed quark doublets, right-handed up quarks and
right-handed leptons, need differentiation among three generations.
Here we assigned the charges $+2$ for the first generation, $+1$ for
the second, and no charge for the third.  This way, top quark Yukawa
coupling is allowed by the flavor charges, but all other Yukawa
couplings are forbidden.  Now suppose the flavor charge is broken by a
small breaking parameter $\epsilon \sim 0.04$ that carries charge
$-1$.  Then all other entries of Yukawa matrices are now allowed but
regulated by powers of the small parameter $\epsilon$.  Then we find
that the ratio of quarks and lepton masses are
\begin{eqnarray}
  \lefteqn{
  m_u : m_c : m_t \sim m_d^2 : m_s^2 : m_b^2} \nonumber \\
  &\sim & m_e^2 : m_\mu^2 : m_\tau^2
  \sim \epsilon^4 : \epsilon^2 : 1.
\end{eqnarray}
Namely, the up quarks are doubly hierarchical than down quarks and
charged leptons, consistent with what we see.  

I'd like to emphasize that the ``anarchy'' is actually a peaceful
ideology, nothing radical.  According to Merriam--Webster dictionary,
anarchy is defined as ``A utopian society of individuals who enjoy
complete freedom without government.''  I'm just saying that neutrinos
work peacefully together to freely mix and abolish hierarchy, without
being forced by any particular structure.  It predicts LMA solution to
the solar neutrino problem, $\sin^2 2\theta_{13}$ must be just below
the current limit, and $O(1)$ CP violating phase.  This is an ideal
scenario for very long-baseline neutrino oscillation experiments,
requiring many countries to be involved.  It is therefore
pro-globalization!

We have seen how new flavor quantum numbers can determine the
structure of masses and mixings among quarks and leptons.  Theorists
of course argue about what the correct charge assignment is.  How we
will know if any of such flavor quantum numbers are actually right?
It will be a long-shot program, needing many new data such as $\sin^2
2\theta_{13}$, solar neutrinos, possible CP violation in the neutrino
sector as well as more details in the quark sector and even charged
leptons (EDM), $B$-physics, Lepton Flavor Violation, even proton
decay.  Because the difference in flavor quantum numbers suppress
flavor mixing, the pattern of flavor violation must be consistent with
assigned flavor quantum numbers.  Details of these flavor-violating
phenomena could eventually tell us what new quantum number assignment
is correct.  It is not clear if the origin of flavor is at the energy
scale accessible by any accelerator experiment; it may be up at the
unification scale.  However, we may still learn enough information to
be able to reconstruct a plausible theory of the origin of flavor,
masses, and mixings.  I call such a program ``archaeology'' in the best
sense of the word.  For example, we can never recreate the conditions
of ancient world.  But by studying fossils, relics, ruins, and
employing multiple techniques, we can come up with satisfactory
plausible theories of what had happened.  Cosmology is by nature an
archaeology.  You can't recreate Big Bang.  But cosmic microwave
background is a wonderfully colorful beautiful dinosaur that tells us
great deal about the history of Universe.  In the same sense, a wealth
of flavor data can point us to the correct theory.

For this purpose, Lepton Flavor Violation (LFV) is a crucial subject.
Now that neutrinos appear to oscillate, {\it i.e.}\/, convert from one
flavor to another, there must be corresponding process among charged
leptons as well.  Unfortunately neutrino masses themselves do not lead
to sizable rates of LFV processes.  However, many extensions of
physics beyond the Standard Model, most notably
supersymmetry,\cite{TASI} tend to give LFV processes at the
interesting levels, such as $\mu \rightarrow e\gamma$, $\mu
\rightarrow e$ conversion, $\tau \rightarrow \mu \gamma$, etc.  The
violation of overall lepton numbers, such as neutrinoless double beta
decay, would be also extremely important.\footnote{See also John Ellis
  at this conference,\cite{Ellis} and papers submitted by Funchal
  Renata Zukanovich, and Bueno et al to this conference.}

\section{Leptogenesis}

One of the primary interest in flavor physics is the origin of the
cosmic baryon asymmetry.  From the Big-Bang Nucleosynthesis, we know
that there is only a tiny asymmetry in the baryon number,
$n_B/n_\gamma \approx 5 \times 10^{-10}$.  In other words, there was
only one excess quark out of ten billion that survived the
annihilation with anti-quarks.  Leptogenesis is a possible origin of
such a small asymmetry using neutrino physics, a possibility that is
gaining popularity now that we seem to see strong evidence for
neutrino mass.

The original baryogenesis theories used grand-unified
theories,\cite{Yoshimura} because grand unification necessarily breaks
the baryon number.  Suppose a GUT-scale particle $X$ decays out of
equilibrium with direct CP violation $B(X \rightarrow q) \neq
B(\bar{X} \rightarrow \bar{q})$.  Then it can create net baryon number
$B \neq 0$ in the final state from the initial state of no baryon
number $B=0$.  It is interesting that such a direct CP violation
indeed had been established in neutral kaon system in this conference
(see R.~Kessler\cite{Kessler} and
L.~Iconomidou-Fayard\cite{Iconomidou} in this proceedings).  However,
the original models preserved $B-L$ and hence did not create net
$B-L$, that turned out be a problem.

The Standard Model actually violates $B$.\cite{anomaly} In the Early
Universe when the temperature was above 250~GeV, there was no Higgs
boson condensate and $W$ and $Z$ bosons were massless (so where all
quarks and leptons).  Therefore $W$ and $Z$ fields were just like
electromagnetic field in the hot plasma and were fluctuating
thermally.  The quarks and leptons move around under the fluctuating
$W$-field background.  To see what they do, we solve the Dirac
equation for fermions coupled to $W$.  There are positive energy
states that are left vacant, and negative energy states that are
filled in the ``vacuum.''  As the $W$-field fluctuates, the energy
levels fluctuate up and down accordingly.  Once in a while, however,
the fluctuation becomes so large that all energy levels are shifted by
one unit.  Then you see that one of the positive energy states is now
occupied.  There is now a particle! This process occurs in the exactly
the same manner for every particle species that couple to $W$, namely
for all left-handed lepton and quark doublets.  This effect is called
the electroweak anomaly.  Therefore the electroweak anomaly changes
(per generation) $\Delta L =1$, and $\Delta q = 1$ for all three
colors, and hence $\Delta B = 1$.  Note that $\Delta (B-L) = 0$; the
electroweak anomaly preserves $B-L$.

\begin{figure}[t]
  \begin{center}
    \includegraphics[width=\columnwidth]{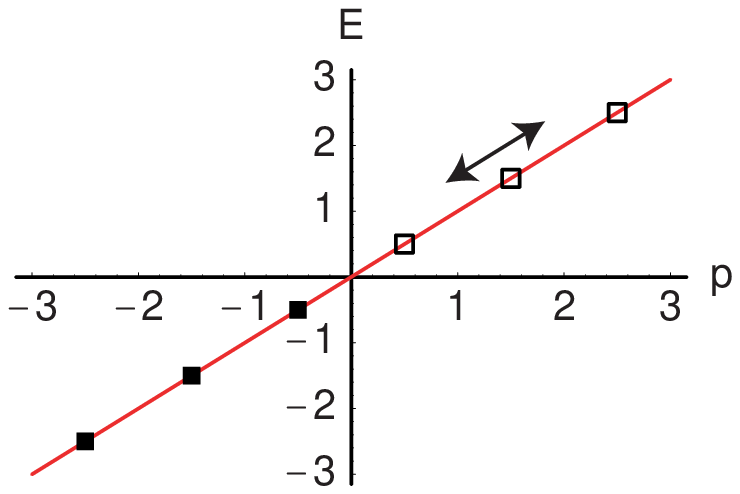}
    \caption[Dirac]{The energy levels of the Dirac equation in the
      presence of fluctuating $W$-field move up and down.  All
      negative energy states are occupied while the positive energy
      states vacant in the ``vacuum'' configuration.}
    \label{fig:Dirac}
    \includegraphics[width=\columnwidth]{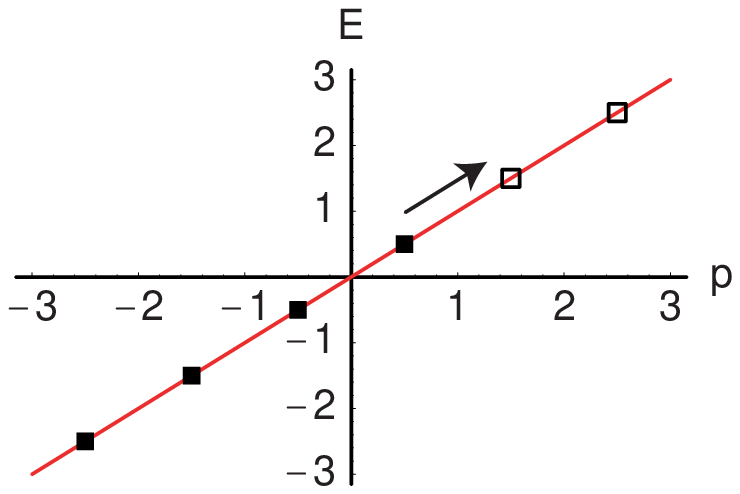}
    \caption[Dirac2]{Once in a while, the fluctuation in the $W$-field
      becomes so large that the energy levels of the Dirac equation in
      the presence of fluctuating $W$-field shift all the way by one
      unit.  Then a positive energy state is occupied and a particle
      is created.}
    \label{fig:Dirac2}
  \end{center}
\end{figure}

Because of this process, the pre-existing $B$ and $L$ are converted to
each other to find the chemical equilibrium at $B \sim 0.35 (B-L)$, $L
\sim -0.65 (B-L)$.\cite{equilibrium} In particular, even if there was
both $B$ and $L$, both of them get washed out if $B-L$ was zero.

Given this problem, there are now two major directions in the
baryogenesis.  One is the electroweak baryogenesis,\cite{KRS} where
you try to generate $B=L$ at the time of the electroweak phase
transition so that they do not get washed out further by the
electroweak anomaly.  The other is the leptogenesis,\cite{FY} where
you try to generate $L\neq 0$ but no $B$ from neutrino physics well
before the electroweak phase transition, and $L$ gets partially
converted o $B$ due to the electroweak anomaly.

The electroweak baryogenesis is not possible in the Standard
Model,\cite{CKN} but is still a possibility in the Minimal
Supersymmetric Standard Model.  However, the model is getting cornered;
the available parameter space is becoming increasingly limited due to
the LEP constraints on chargino, scalar top quark and Higgs
boson.\cite{Carena,Cline} We are supposed to find a right-handed
scalar top quark, charginos ``soon'' with a large CP violation in the
chargino sector.  There is possibly a detectable consequence in
$B$-physics as well.\cite{EWbaryogenesis2}

\begin{figure}[t]
  \begin{center}
    \includegraphics[width=\columnwidth]{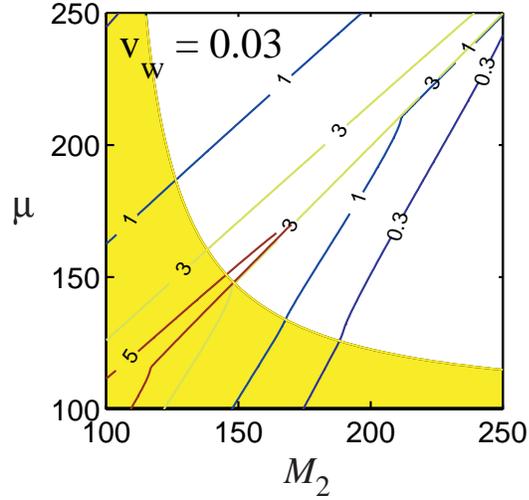}
    \caption[EWbaryogenesis]{Constraint on the MSSM chargino parameter
      space in electroweak baryogenesis.\cite{Cline}  To generate
      $\eta = 5 \times 10^{-10}$, the parameters must lie inside the
      contour labeled ``5.''  It implies light charginos.  Shaded
      region is excluded by LEP.}
    \label{fig:EWbaryogenesis}
  \end{center}
\end{figure}

\begin{figure}[t]
  \begin{center}
    \includegraphics[width=\columnwidth]{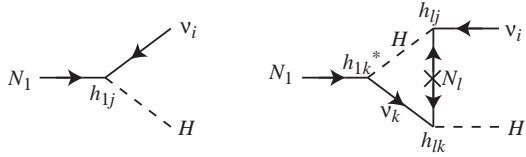}
    \caption[Ndecay]{The tree-level and one-loop diagrams of
      right-handed neutrino decay into leptons and Higgs.  The
      absorptive part in the one-loop diagram together with
      CP-violating phases in the Yukawa couplings leads to the direct
      CP violation $\Gamma (N_1 \rightarrow l H) \neq \Gamma (N_1
      \rightarrow \bar{l} H)$.}
    \label{fig:Ndecay}
  \end{center}
\end{figure}

In leptogenesis, you generate $L \neq 0$ first.  Then $L$ gets
partially converted to $B$ by the electroweak anomaly.  The question
then is how you generate $L \neq 0$.  In the original
proposal,\cite{FY} it was done by the decay of a right-handed neutrino
(say $N_1$), present in the seesaw mechanism, with a direct CP
violation.  At the tree-level, a right-handed neutrino decays equally
into $l+H$ and $\bar{l}+H^*$.  At the one-loop level, however, the
interference between diagrams shown in Fig.~\ref{fig:Ndecay} cause a
difference in the decay rates of a right-handed neutrino into leptons
and anti-leptons proportional to $\Im (h_{1j} h_{1k} h_{lk}^*
h_{lj}^*)$.  Much more details had been worked out in the light of
recent neutrino oscillation data and it had been shown that a
right-handed neutrino of about $10^{10}$~GeV can well account for the
cosmic baryon asymmetry from its out-of-equilibrium
decay.\cite{Buchmuller} There is some tension in the supersymmetric
version because of cosmological problems caused by the gravitino.  But
it can be circumvented for instance using the superpartner of
right-handed neutrino that can have a coherent oscillation after the
inflation.\cite{Hamachan} Leptogenesis can work.

Can we prove leptogenesis experimentally?  Lay Nam Chang, John Ellis,
Belen Gavela, Boris Kayser, and myself got together at Snowmass and
discussed this question.  The short answer is unfortunately no.  There
are additional CP violating phases in the heavy right-handed neutrino
sector that cannot be seen by studying the light left-handed
neutrinos.  For example, even two-generation seesaw mechanism is
enough to have CP violation that can potentially produce lepton
asymmetry, unlike the minimum of three-generations for CP violation in
neutrino oscillation.  However, we decided that if we will see (1)
electroweak baryogenesis ruled out, (2) lepton-number violation {\it
  e.g.}\/ in neutrinoless double beta decay, and (3) CP violation in
the neutrino sector {\it e.g.}\/, in very long-baseline neutrino
oscillation experiment, we will probably believe it based on these
``archaeological'' evidences.  

Also, there may be additional fossils of the leptogenesis that depend
on more details of the model.  For instance, in the supersymmetric
version with coherent right-handed sneutrino,\cite{Hamachan} a small
isocurvature component in density perturbation is created because the
overall energy density and the baryon asymmetry may have independent
fluctuations.  If you believe in a certain scenario of supersymmetry
breaking, low-energy lepton-flavor violation can carry information
about CP violation in the right-handed neutrino sector.\cite{Hisano}
Combining the above three general requirements, lepton-flavor
violation and cosmological tests, the case for leptogenesis can become
even stronger.

\section{Future}

Even though dramatic progress had been made, many questions remain.
The important aspect of neutrino physics is that we don't stop here.
It is quite a healthy field with many studies done jointly by
theorists and experimentalists looking forward.  Here I will discuss
what is coming in the near future.\footnote{Plots are shown only for
  experiments that start data taking by the next Lepton Photon
  conference.}

First of all, the oscillation signal from LSND will be verified or
refuted at high confidence levels by Mini-BooNE experiment.  It will
start taking data in 2002.  The result will mark a major branch point
in the development of neutrino physics: do we need a sterile neutrino?

\begin{figure}[t]
  \begin{center}
    \includegraphics[width=\columnwidth]{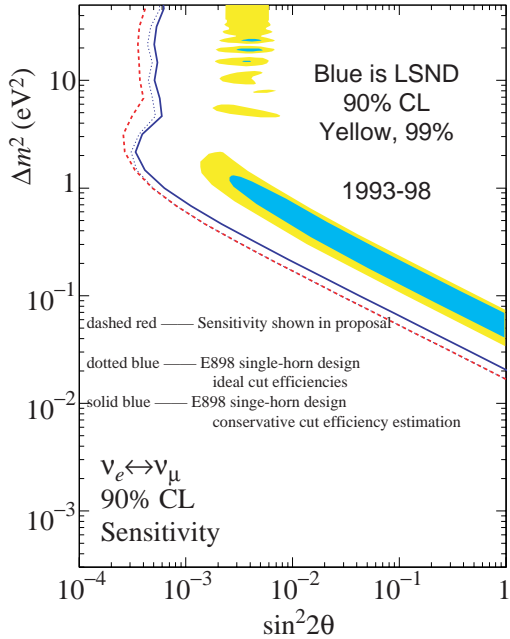}
    \caption[Mini-BooNE]{MiniBooNE expected 90\% confidence level
      sensitivity limits.\cite{Mini-BooNE}}
    \label{fig:Mini-BooNE}
  \end{center}
\end{figure}

There is a series of experiments aimed at the oscillation signal in
the atmospheric neutrinos after K2K.  MINOS, a long-baseline
oscillation experiment from Fermilab to Sudan, Minnesota, will start
in 2004, and will determine $\Delta m^2_{23}$ precisely.  This will
provide crucial input to design very long-baseline neutrino
oscillation experiments of later generations.  OPERA and ICARUS in
Gran Sasso, using the CGNS beam from CERN, will look for $\tau$
appearance in the $\nu_\mu$ beam for the same $\Delta m^2_{23}$.  All
of these experiments extend reach in $\sin^2 2\theta_{13}$ as well.
MONOLITH, if approved, will study atmospheric neutrinos using iron
calorimetry and verify the oscillation dip.

\begin{figure}[t]
  \begin{center}
    \includegraphics[width=\columnwidth]{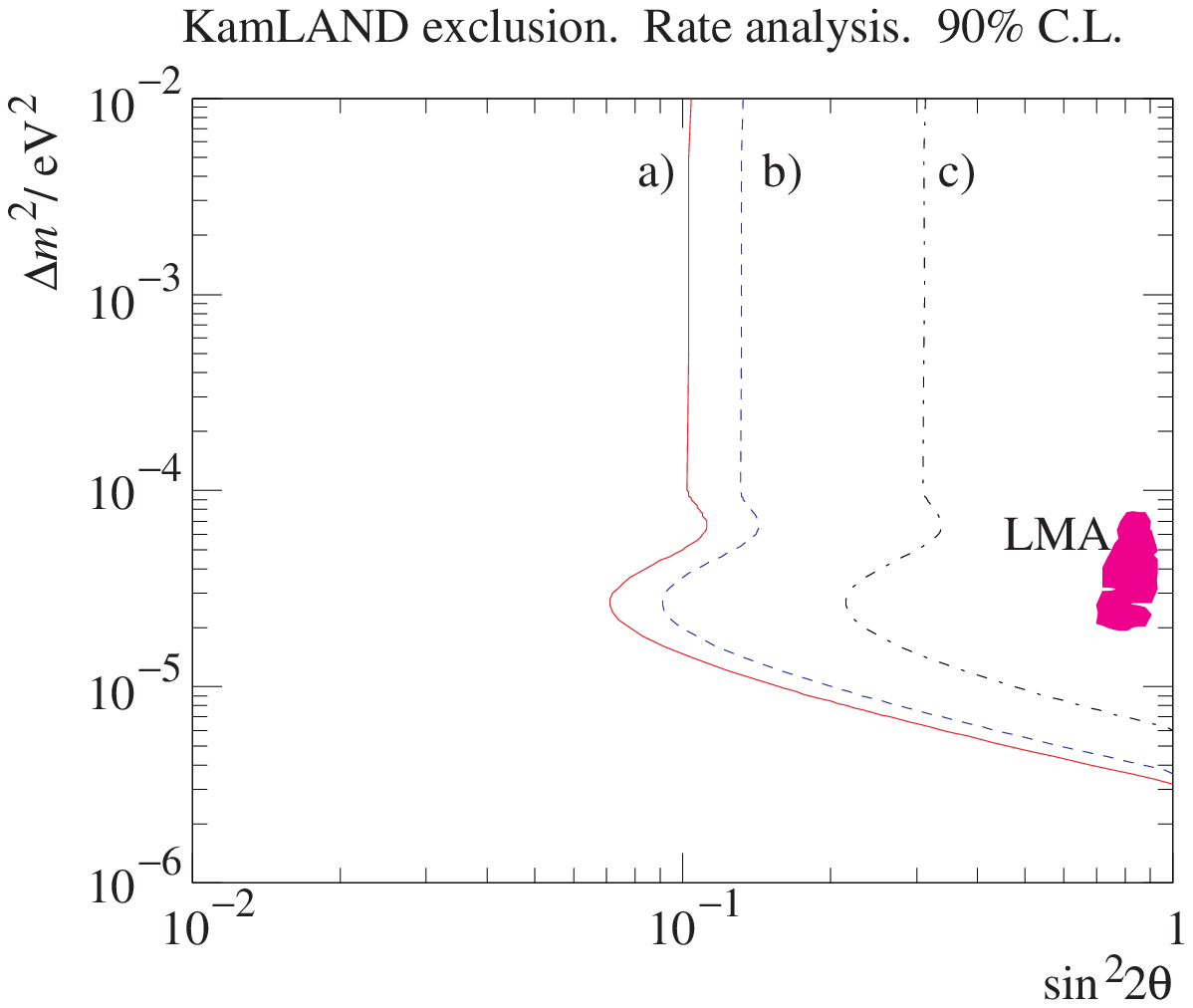}
    \caption[KamLAND]{KamLAND expected 90\% confidence level
      sensitivity limits.\cite{KamLAND}}
    \label{fig:KamLAND}
    \includegraphics[width=\columnwidth]{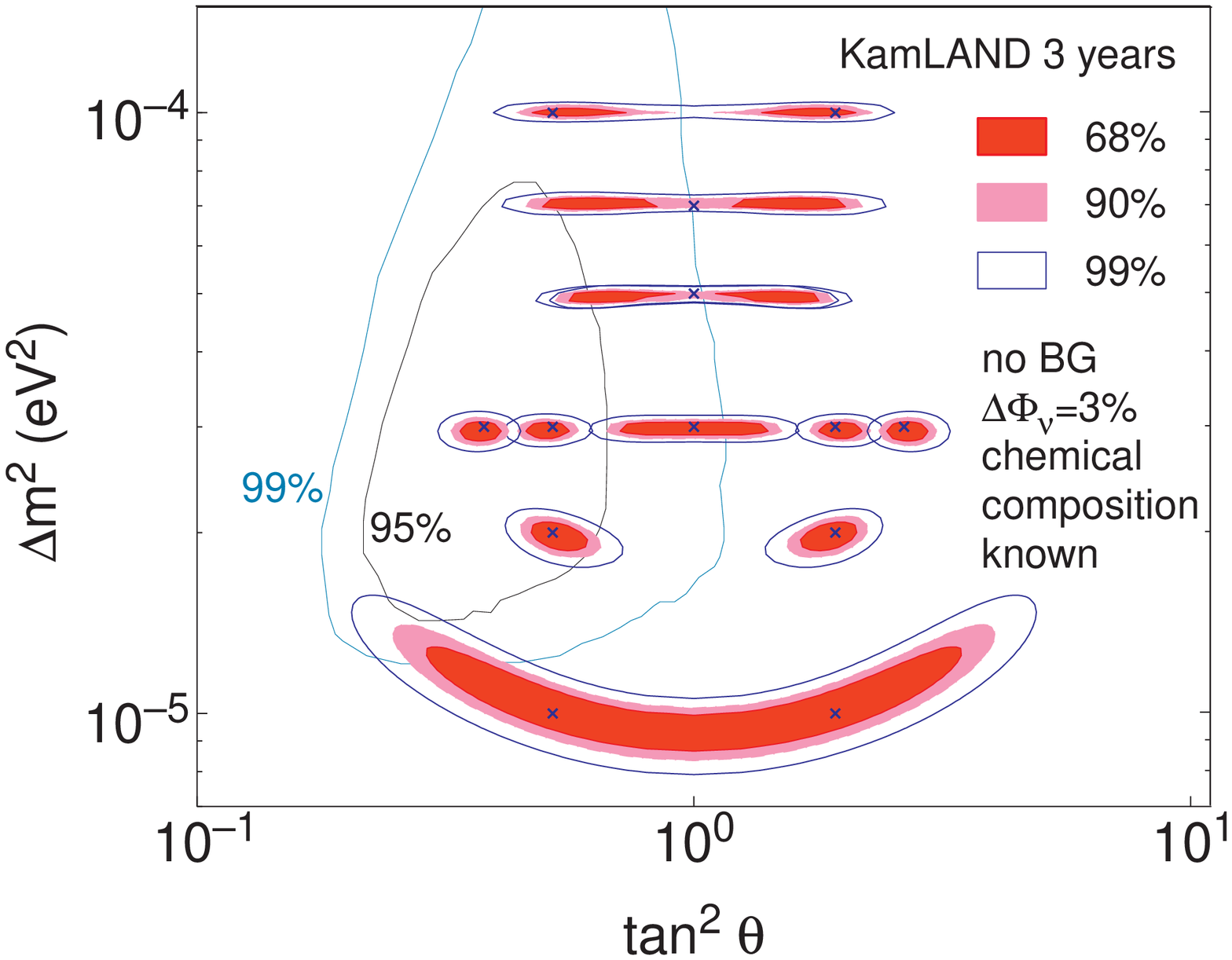}
    \caption[KamLANDmeas]{Measurement of oscillation parameters at
      KamLAND.\cite{APHM}}
  \end{center}
\end{figure}

On the solar neutrino oscillation signal, KamLAND will be the first
terrestrial experiment to attack this problem.  Construction has
completed in 2001.  Using reactor neutrinos over the baseline of about
175~km, it will verify or exclude currently-favored LMA solution to
the solar neutrino problem.  If signal will be found, it will
determine oscillation parameters quite
well.\cite{KamLANDBarger,APHM,KamLANDAndre} If this turns out to be
the case, it will be a dream scenario for neutrino oscillation
physics.  This is because $\Delta m^2_{\rm solar}$ is within the
sensitivity of terrestrial-scale very long baseline experiments, making
CP violation in neutrino oscillation a possible target.  For example,
the difference in neutrino and anti-neutrino oscillation rates is
given by
\begin{eqnarray}
  \lefteqn{
  P(\nu_e \rightarrow \nu_\mu) - P(\bar{\nu}_e \rightarrow
  \bar{\nu}_\mu) } \nonumber \\
  & &= 16 s_{12} c_{12} s_{13} c_{13}^2 s_{23} c_{23}
  \sin \delta \nonumber \\
  & & \sin \frac{\Delta m_{12}^2}{4E}L
  \sin \frac{\Delta m_{13}^2}{4E}L
  \sin \frac{\Delta m_{23}^2}{4E}L 
\end{eqnarray}
using the notation of the MNS matrix in Eq.~(\ref{eq:UMNS}).  $\delta$
is the CP-violating phase.  Clearly, $\Delta m^2_{12}$ has to be
sizable in order for the difference not to vanish.  At the same time,
large $s_{13}$ is preferred.  On the other hand, if LMA will be
excluded, study of low-energy solar neutrinos will be crucial.
KamLAND will be another major branch point.

\begin{table*}[t]
\caption[E1]{Physics Sensitivity for Current Superbeam Proposals.\cite{E1}} 
\label{tab:sb1} 
\begin{tabular}{llllll}
\hline\hline
Name & Years of Running & kton & 
$\sin^2 2\theta_{13}$ sen- & CP Phase $\delta$ & $E_\nu$ \\ 
  &   &   &  sitivity (3$\sigma$) & sensitivity (3$\sigma$) & (GeV) 
\\ \hline
JHF to SuperK & 5 years $\nu$ & 50 & 0.016 & none & 0.7 \\ \hline
SJHF to HyperK & 2 years $\nu$, 6 years $\bar\nu$ & 1000& 0.0025 &
$>15^\circ$  & 0.7 \\ \hline 
CERN to UNO & 2 years $\nu$, 10 years $\bar\nu$ & 400 & 0.0025 & $>40^\circ$ 
& 0.3 \\
\hline
\end{tabular} 
\end{table*} 

There are already proposals to extend reach in $\sin^2 2\theta_{13}$
as well as possibly detect CP violation, often called neutrino
superbeam experiments.  One possibility is to build a neutrino beam
line using 50~GeV protons from JHF (Japan Hadron Facility) under
construction, and aim the beam at SuperKamiokande.  A possible upgrade
of SuperKamiokande to an even bigger HyperKamiokande (?) together with
more intense neutrino beam is also being discussed.  On the other
hand, CERN is discussing the Super Proton Linac (SPL), a GeV proton
accelerator using LEP superconducting cavities.  It can be aimed at a
water Cherenkov detector at a modest distance to study possible CP
violation thanks to its low energy.  See Table~\ref{tab:sb1} for their
sensitivity.

At Snowmass, a case was made that a higher energy superbeam with a
longer baseline will be beneficial.\cite{E1} For example, the matter
effect can discriminate between two possible mass spectra in
Fig.~\ref{fig:2+1}, and a longer baseline is needed for this purpose.

\begin{figure}[t]
  \begin{center}
    \rotatebox{-90}{
    \includegraphics[height=\columnwidth]{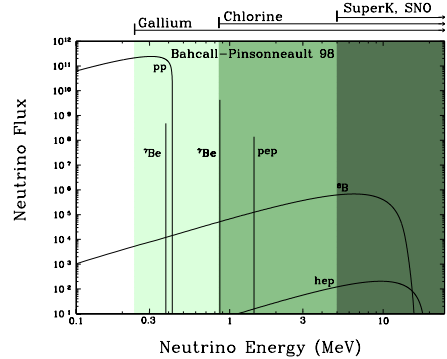}
    }
    \caption[solarneutrinospectrum]{The spectrum of solar neutrinos.\cite{BP}}
    \label{fig:solarneutrinospectrum}
  \end{center}
\end{figure}

If KamLAND excludes the LMA solution, the study of low-energy solar
neutrinos will be crucial.  The predicted spectrum of solar neutrinos
is shown in Fig.~\ref{fig:solarneutrinospectrum}.  Only real-time
experiments had been done so far using $^8$B neutrinos.  

\begin{figure}[t]
  \begin{center}
    \includegraphics[width=\columnwidth]{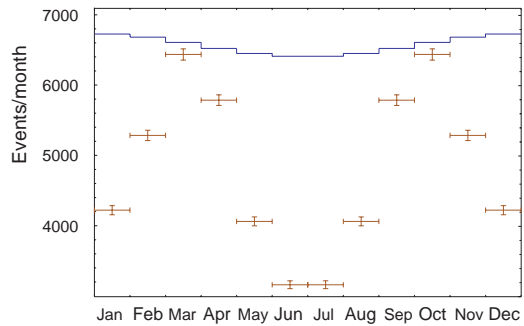}
    \caption[vacuumosc]{Illustration of the effect of 
      vacuum oscillations on the shape of the seasonal variation of
      the solar neutrino data.  The points with statistical error bars
      represent the number of events/month expected at Borexino after
      3 years of running for $\Delta m^2 = 3 \times 10^{-10}$~eV$^2$,
      $\sin ^2 2\theta = 1$.\cite{seasonal}}
    \label{fig:vacuumosc}
  \end{center}
\end{figure}

\begin{figure}[t]
  \begin{center}
    \includegraphics[width=\columnwidth]{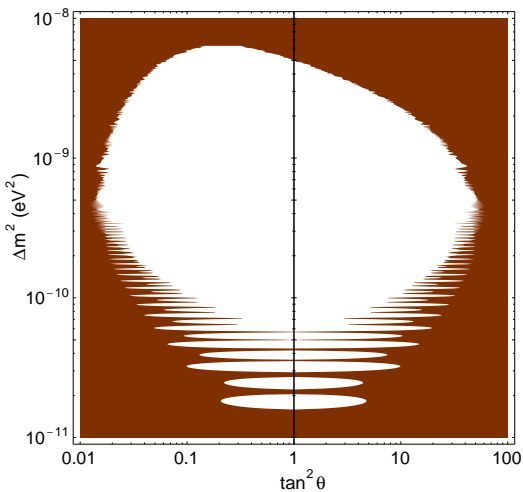}
    \caption[vacuum]{The parameter space Borexino is sensitive to by
      looking for anomalous seasonal variation in the event rate.}
    \label{fig:vacuum}
  \end{center}
\end{figure}

Borexino will be the first real-time experiment to detect lower-energy
solar neutrinos from the $^7$Be line.  They expect data taking
starting in 2002.  What is so crucial about the $^7$Be neutrinos is
that it is mono-energetic (with some broadening due to thermal
collisions in the Sun).  The VAC solution to the solar neutrino
problem can be studied by looking for anomalous seasonal variation.
The distance between the Sun and the Earth is not constant; the Sun is
closer in the winter while farther in the summer in the northern
hemisphere.  This seasonal change in the distance causes a change in
the oscillation probability for this solution and the event rate would
be modulated beyond the trivial factor of $1/r^2$; see
Fig.~\ref{fig:vacuumosc}.  Borexino can look for such an anomalous
seasonal variation.  The sensitivity region is shown in
Fig.~\ref{fig:vacuum}.  At the low $\Delta m^2$ region, the region is
reflection symmetric. However for larger values of $\Delta m^2$, the
region is asymmetric showing the impact of matter effect.  The fact
that matter effect is important even for the VAC solution had not been
realized until recently.\cite{Alex} This parameter region was named
``quasi-vacuum''.\cite{quasivacuum}

The LOW solution would lead to a large Earth matter effect for $^7$Be
neutrinos.\cite{daynight,daynight2}  Again once found, the zenith
angle dependence of the event rate depends sensitively on the
oscillation parameters and can be determined quite well.

\begin{figure}[t]
  \begin{center}
    \includegraphics[width=\columnwidth]{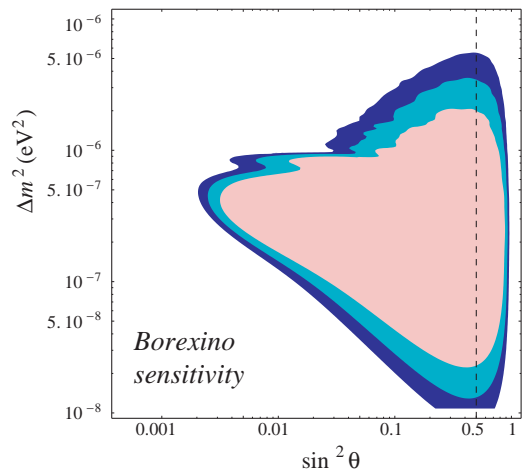}
    \caption[LOW]{95\% (darkest), 3$\sigma$ (dark), and
      5$\sigma$ (light) sensitivity confidence level (C.L.) contours
      for three years of Borexino running.\cite{daynight} The LOW
      solution is completely covered.}
    \label{fig:LOW}
  \end{center}
\end{figure}

If the SMA solution turns out to be true, SNO may eventually see a
distortion in the charged-current spectrum.  However, the
determination of parameters would require the study of $pp$ neutrinos.
The point is that there is a sharp falloff in the survival parameter
in the $pp$ spectrum.  This is because the level-crossing does not
occur for low energy for which the matter effect does not overcome the
mass splitting $\frac{\Delta m^2}{2E} > \sqrt{2} G_F n_e(0)$.  The
location of the falloff determines $\Delta m^2$.  On the other hand,
the slow rise at higher energy depends both on the mixing angle and
the mass-squared difference.

\begin{figure}[t]
  \begin{center}
    \includegraphics[width=\columnwidth]{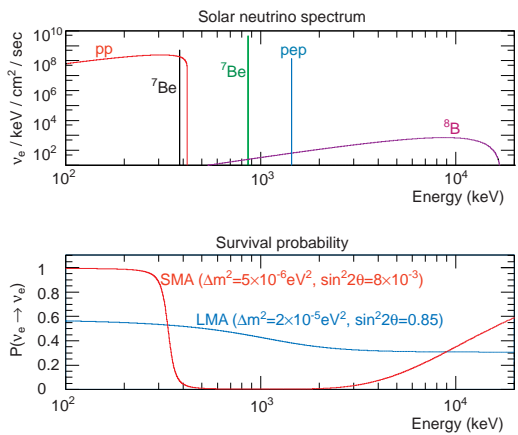}
    \caption[SMA]{The survival probability as a function of neutrino
      energy.  For SMA, there is a sharp falloff in the survival
      probability. }
    \label{fig:SMA}
  \end{center}
\end{figure}

GALLEX, SAGE, and now GNO experiments had detected $pp$ neutrinos, but
they cannot measure neutrino energy spectrum and hence cannot separate
$pp$ neutrinos from other solar neutrinos.  There had been many
proposals to study $pp$ neutrinos.  Some of them use neutrino-electron
elastic scattering, sensitive to both charged- and neutral-current
interactions (gaseous Helium TPC, HERON with superfluid He, liquid Xe,
GENIUS using solid Ge), while others use charged-current reactions on
nuclei (LENS, both Yb and In versions, and Moon on Mo).  I hope that
some of them will work in the end and provide us crucial information.

\section{Conclusions}

Neutrino physics is going through a revolution right now.  We had
learned a lot already, and will learn a lot more, especially on solar
neutrinos.  It provides an unambiguous evidence for physics beyond the
minimal Standard Model.

Given strong evidences for neutrino mass, leptogenesis has gained
momentum as the possible origin of cosmic baryon asymmetry.
Establishing lepton-number violation would be crucial, while seeing CP
violation in neutrino oscillation would boost the credibility.
Neutrino superbeams and eventually neutrino factory could play
essential role in this respect if the currently-favored LMA solution
turns out to be correct.

We may not be lucky enough to test directly leptogenesis and models of
neutrino masses, mixings, and flavor in general.  But we can collect
more ``fossils'' to gain insight into flavor physics at high energies,
such as lepton flavor violation, combination of quark flavor physics,
and even proton decay.  Data may eventually point to new flavor
quantum numbers that control the masses and mixings.

\section*{Acknowledgments}
This work was supported in part by the DOE Contract DE-AC03-76SF00098
and in part by the NSF grant PHY-0098840.

\end{document}